\documentclass[journal]{IEEEtran}
\ifCLASSINFOpdf
\else
\fi
\usepackage[table]{xcolor}
\usepackage{amssymb}
\usepackage{multirow}
\usepackage{graphicx}
\usepackage{amsmath}
\usepackage{booktabs}
\usepackage{multirow}
\usepackage{siunitx}
\begin{document}
% \renewcommand{\baselinestretch}{1.5}
% \onecolumn


%\title{One's accepted, two's dropped: three high performance CNN architectures for frame-level co-channel speech detection}
\title{Single-channel speech separation using Soft-minimum Permutation Invariant Training}


\author{Midia Yousefi,~\IEEEmembership{Student Member,~IEEE,}
        and~John Hansen,~\IEEEmembership{Fellow,~IEEE}% <-this % stops a space
\thanks{M. Yousefi and J. H. L Hansen are with the Center for Robust Speech Systems, Erik Jonsson School of Engineering and Computer Science, The University of Texas at Dallas, Richardson, TX 75080 USA (e-mail:, midia.yousefi@
utdallas.edu; john.hansen@utdallas.edu).}% <-this % stops a space
\thanks{Manuscript received August, 2021.}}


\markboth{IEEE/ACM TRANSACTIONS ON AUDIO, SPEECH, AND LANGUAGE PROCESSING,~Vol.~14, No.~8, August~2021}%
{Shell \MakeLowercase{\textit{et al.}}: Bare Demo of IEEEtran.cls for IEEE Journals}



\maketitle

\begin{abstract}

The goal of speech separation is to extract multiple speech sources from a single microphone recording.  Recently, with the advancement of deep learning and availability of large datasets, speech separation has been formulated as a supervised learning problem. These approaches aim to learn discriminative patterns of speech, speakers, and background noise using a supervised learning algorithm, typically a deep neural network. A long-lasting problem in supervised speech separation is finding the correct label for each separated speech signal, referred to as label permutation ambiguity. Permutation ambiguity refers to the problem of determining the output-label assignment between the separated sources and the available single-speaker speech labels. Finding the best output-label assignment is required for calculation of separation error, which is later used for updating parameters of the model. Recently, Permutation Invariant Training (PIT) has been shown to be a promising solution in handling the label ambiguity problem. However, the overconfident choice of the output-label assignment by PIT results in a sub-optimal trained model. In this work, we propose a probabilistic optimization framework to address the inefficiency of PIT in finding the best output-label assignment. Our proposed method entitled trainable Soft-minimum PIT is then employed on the same Long-Short Term Memory (LSTM) architecture used in Permutation Invariant Training (PIT) speech separation method. The results of our experiments show that the proposed method outperforms conventional PIT speech separation significantly (p-value $ < 0.01$) by +1dB in Signal to Distortion Ratio (SDR) and  +1.5dB in Signal to Interference Ratio (SIR).

\end{abstract}

\begin{IEEEkeywords}
source separation, speech separation, cocktail party, probabilistic permutation invariant training, PIT, prob PIT, soft-minimum PIT
\end{IEEEkeywords}

% \fi
\IEEEpeerreviewmaketitle

\section{Introduction}

Extracting the underlying sources from a signal mixture is a general problem in many applications. A classical example for such an application is to recognize or isolate what is being said by an individual speaker in a cocktail-party scenario in which multiple speakers are talking simultaneously \cite{yousefi2020block}. The auditory system of the human brain encounters two main challenges in the cocktail party scenario. First, it carries out sound segregation, which is the act of deriving properties of the individual sources from the mixture \cite{carlyon1992psychophysics}. Second, it can switch attention between different sources when following distinct conversations \cite{shinn2008object,koch2011switching}. Humans can accomplish this in part due to bilateral hearing, as well as learned effective neural decoding in the auditory cortex.

\begin{figure*}
\begin{center}
\includegraphics[width=\linewidth,height=5.5cm]{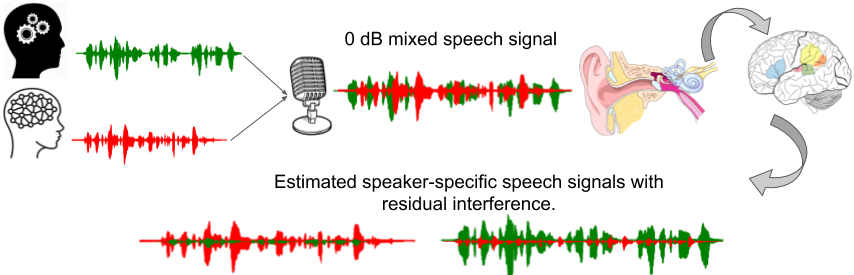}
\caption{Speech separation performed by human auditory system. The separated speech sources still contain residual speech from interfering talker. }
\label{fig:cock}
\vspace{-0.5cm}  
\end{center}
\end{figure*}

However, as shown in Fig. \ref{fig:cock}, listening and following one speaker in the presence of competing speakers is an easy task for the majority of people, a remarkable ability usually taken for granted. While extensive research has explored speaker recognition by machines \cite{stoter2018countnet}, the current task requires expanded knowledge and capabilities.  However, even for humans with normal hearing abilities, the capacity of the human auditory system to extract and separate simultaneous sources out of a mixture is severely compromised \cite{stoter2018countnet,bronkhorst2015cocktail, yousefi2021speaker}. As reported in \cite{kawashima2015perceptual}, humans are capable of detecting up to three simultaneous active speakers without using spatial information of the input mixture. Thus, solving the cocktail party problem for mixtures with more than three concurrent active speakers is a very challenging task in which even humans may not be able to address \cite{kashino1996one}.

%Concerning the cocktail party challenge, one critical question rises; ``Is it possible to build a machine with the same auditory capabilities as the humans, which is to extract and isolate an individual source at a time?''. The answer to this question lies in engineered signal processing disciplines and machine learning techniques. For many years, researches have successfully developed tools and methods to mimic the brains auditory system. Almost all the state-of-the-art speech technologies such as Automatic Speech Recognition (ASR), speaker diarization, speaker identification and speech synthesis systems operate perfectly when the input is a clean single speaker signal. In contrast, these systems perform much worse in the real world scenarios especially in existence of an interfering talker \cite{weng2015deep,hershey2010super,boakye2008overlapped}.

The cocktail party problem can be viewed as addressing blind source separation \cite{qian2018past}, which is the task of recovering a set of independent sources when only their mixtures with unknown coefficients are available. Source separation can be considered as the combination of many active research threads such as: speech enhancement \cite{vincent2018audio}, speech separation \cite{makino2007blind, yousefi2021real}, waveform preserving estimation \cite{tong1993waveform, sidiropoulos1998deterministic}, music separation \cite{ozerov2007adaptation,luo2017deep}, etc. Each of these research sub-communities make specific assumptions on the structure and properties of the active sources in the mixture, which results in problem-specific solutions to the cocktail party problem. Given the central role of overlapping speech detection and separation in the cocktail party scenario, in this study, we focus on single-channel speaker-independent speech separation. 

%Solving this problem requires developing computer algorithms that are capable of mimicking humans auditory systems in the challenging scenarios that will enable a truly free conversation between human and machines.

The pioneering work of separating audio signals from a mixture was represented by Bregman in \cite{bregman1990auditory}. He noted that the auditory system performs an Auditory Scene Analysis (ASA) on the mixture signal entering the ear. ASA is executed in two steps: \emph{(i)} the acoustic signal is decomposed into a number of sensory components, and \emph{(ii)}  the components that are likely from the same source are combined into a single stream. Based on this study, Computational Auditory Scene Analysis (CASA) was proposed in \cite{brown1994computational}. CASA attempts to model different parts of the biological auditory system that encomposes outer ear, middle ear and inner ear. In this approach, the signal is passed through a set of filterbanks that mimics the sound transduction performed by the inner hair cells. Next, for each filterbank output, periodicity, frequency transition, oneset and offset are calculated. Using pitch tracking techniques and labeling T-F bins, CASA groups the voice and unvoiced segments likely belonging to the same source. Finally, each waveform is reconstructed based on the source-specific segments \cite{brown1994computational}. 

Independent Component Analysis (ICA) is another main technique introduced to address source separation \cite{comon1994independent}. The strength of ICA as a separation tool resides in a realistic assumption that different physical processes generate unrelated signals. Therefore, when given a mixture of multiple speech sources, ICA identifies those unrelated signals which are voice traits of different speakers \cite{comon2010handbook}. In ICA, data vectors are represented using weighted a linear combination of basis functions. Higher order statistics are needed to derive the independent coefficients of each basis. ICA removes not only correlation but also higher order dependence between the estimated bases, which has contributed to its success in extracting individual sources \cite{lee1998independent}. Several years later, a powerful decomposition method called Non-Negative Matrix Factorization (NMF) was introduced, which has been very effective in modeling latent structure of data \cite{lee1999learning}.  NMF finds the parts-based representation of non-negative data through matrix decomposition, and therefore it is capable of extracting  underlying  speech sources  from  a  mixture \cite{lee2001algorithms}. A number of techniques have been developed based on NMF \cite{hoyer2004non,ding2005equivalence}. Sparse NMF and Convolutive NMF (CNMF) are among the most popular \cite{yousefi2016supervised}. CNMF outperforms NMF by modeling the temporal continuity of the speech signal in a time span of several frames \cite{o2006convolutive,smaragdis2006convolutive}. Although both ICA and NMF are supervised machine learning approaches and can learn useful patterns from the input data, the linear structure of the trained model in both ICA and NMF prevents it from learning complex structure within the speech signal. Thus, non-linear machine learning approaches such as Deep Neural Networks have been of interest. 

Recently, learning-based approaches have boosted the performance of speech separation dramatically \cite{wang2018supervised, hershey2016deep, yousefi2021system}. Deep Clustering (DPCL) \cite{hershey2016deep} was among the first approaches that made significant progress in extracting speech signals out of a mixture without prior information concerning the number of speakers. DPCL converts the mixture speech into an embedding space using an Recurrent Neural Network (RNN), with hope that T-F bins belonging to the same speaker establish a cluster in the embedding space. K-means clustering is used in the embedding space to identify these clusters. Finally, another network is trained based on T-F bins grouped in each cluster to estimate source-specific masks in order to recover the individual speech signals from the mixture \cite{hershey2016deep}. Another related technique called Deep Attractor Network (DANet) was introduced in \cite{chen2017deep}. Similar to DPCL, DANet projects the T-F bins of the mixture into an embedding space. DANet uses Exoectation-Maximization (EM) to represent each  speaker in the embedding space using a vector called an attarctor point such that the T-F bins belonging to that speaker are pulled toward the corresponding attractor point.  Finally, the speech signals are estimated based on the grouped T-F bins around each attarctor point \cite{luo2018speaker}. 

Permutation Invariant Training (PIT) \cite{yu2017permutation,kolbaek2017multitalker} is another effective solution which performs separation in two steps: first, it trains a neural network to separate the specific speech sources, and second, it finds the best output-label assignment to minimize the separation error. However, since the network generates unreliable outputs in the initial steps of training, the costs of different output-label permutations are close. The inefficiency of PIT in addressing permutation ambiguity has been considered in our previous study \cite{yousefi2019probabilistic}. Therefore, we proposed Probabilistic PIT (Prob-PIT) which defines a log-likelihood function based on separation errors of all possible permutations. Unlike conventional PIT that uses one output-label permutation with the minimum cost, Prob-PIT uses all permutations by employing the soft-minimum function \cite{yousefi2019probabilistic}. The effectiveness of Prob-PIT is achieved only when the parameters of the log-likelihood cost function are tuned well for each dataset. This can be very tedious and may require extensive time and computational resources to find the best hyperparameter value. To address this issue in Prob-PIT, in this study, we build on our previous work \cite{yousefi2019probabilistic} and propose a novel trainable Probabilistic PIT which we call Soft-minimum PIT to resolve the label permutation ambiguity challenge without requiring manual tuning of the parameters of the log-likelihood cost function. The contributions of this study are threefold:

\begin{itemize}
    \item Proposing a novel trainable Probabilistic Permutation Invariant Training framework called soft-minimum PIT for single-channel supervised speech separation.
    
    \item Comparing the results of both training and tuning of the hyperparameters of the log-likelihood cost function.
  
    \item Comparing the results of our proposed system with the conventional PIT.
  
\end{itemize}

The remainder of the paper is organized as follows. We present the problem formulation, generating overlapping speech mixtures, and extracting spectral features in Sec. \ref{sec:problem}. Details of the proposed trainable Prob-PIT are explained in Sec. \ref{sec:net}. We report on the experimental procedures and results in Sec. \ref{sec:exp}. The results are discussed in Sec. \ref{sec:dis}, and finally the conclusion is presented in the last section.

\begin{figure*}
\begin{center}
\includegraphics[width=\linewidth]{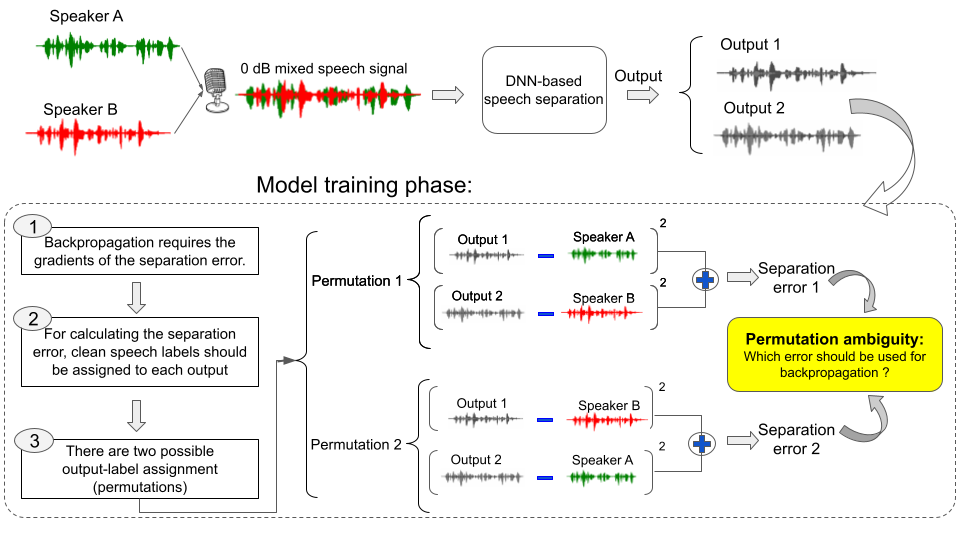}
\caption{Single-channel speaker-independent speech separation pipeline. Once the speaker-specific speech signals have been recovered at the output of the separation system, they should be assigned to their corresponding clean speech versions i.e. speaker A or speaker B for separation loss calculation. The permutation ambiguity caused by different possible output-label assignments leads to different separation errors causing gradient conflict in the training phase. }
\label{fig:perm}
\vspace{-0.5cm}  
\end{center}
\end{figure*}

\section{Problem formulation}
\label{sec:problem}

Speech separation is extremely challenging under the single-microphone speaker-independent scenario, where no prior speaker information is available during evaluation. In supervised approaches, speech separation is formulated as a linear combination of single-speaker speech signals:
\begin{equation}
    y[n] = \sum_{s=1}^{S} x_s[n]
    \label{eq:sep}
\end{equation}
in which $S$ is the total number of speakers in the mixture signal $y$, and $x_{s}$ is the speech signal corresponding to speaker $s$. While  the number of total speakers $S$ is assumed as available prior information, the goal of speech separation is to estimate the speaker-specific speech signals $x_{s}$ from the mixture $y$. In this study, we consider the case of two-talker mixed speech separation. Therefore, Eq. \ref{eq:sep} is modified as:

\begin{equation}
    y[n] = x_1[n] + x_2[n]
    \label{eq:2-sep}
\end{equation}
with $x_1[n]$ and $x_2[n]$ represent the target and interfering speakers respectively. Speech separation is generally performed in the frequency domain by transforming the signals using the Short-Time Fourier Transform (STFT). The main reason for this choice is that the speech structure such as harmonics, formants, and energy densities of are better represented in this domain. Therefore, speech separation is formulated as the task of recovering (STFT) of the source signals $X_s(t, f)$ for each time frame $t$ and frequency bin $f$, given the mixed speech. However, since estimating phase information of this STFT representation is still an open problem \cite{williamson2016complex}, the phase information is acquired from the overlapping speech signal, and the separation is simplified to the task of estimating the magnitude spectra of the speaker-specific speech signals $X_1$ and $X_2$ from the mixture $Y$ as:

\begin{equation}
    Y(f,t) = X_1(f,t) + X_2(f,t),
    \label{eq:mag-sep}
\end{equation}
where $Y$ and $X$ represent the magnitude spectra of $y$ and $x$. The estimation of magnitude spectra of speech is usually achieved by training a model using supervised learning techniques. However, due to label permutation ambiguity, training a robust model for speech separation is challenging. Permutation ambiguity as depicted in Fig. \ref{fig:perm}, happens only in the model training stage and affects system performance in the testing phase. In Fig. \ref{fig:perm}, a single-channel mixed speech signal is processed by a DNN-based separation system. The goal of this separation is to extract speech signals corresponding to speakers $A$ and $B$ from the input mixture. Since there are only two speech sources in this mixture, the separation system has two outputs named $o_1$ and $o_2$. Each output contains a separated speech waveform corresponding to one of the speakers in the mixed speech. If the separation system is used for test, estimated  speech signals at output 1 $o_1$ and output 2 $o_2$ are the separated final streams. However, in training, additional processing steps are required for updating the model parameters in each epoch. In supervised learning, updating parameters of the model is accomplished by comparing model's output with the desired label, which in this application is the single-speaker speech waveform. The more the model output is similar to the desired label, the less the model's parameters are updated.  

One important component in training, is backpropagation of the separation error which is the process of calculating the gradient of the error function with respect to the neural network's weights. Therefore, the effectiveness of backpropagation is highly dependent on the correct and precise value of the separation error. At this point, this question rises: "Which single-speaker speech waveform should be used as the desired form of each output?". Effectively, we should find the correct order between the outputs and the labels. There are two possible solutions:  

\vspace{0.1cm}
\textbf{Scenario 1:}
\begin{itemize}
    \item Output1: is the estimated speech related to Speaker A.
    \item Output2: is the estimated speech related to Speaker B.
\end{itemize}

\vspace{0.1cm}
\textbf{Scenario 2:}
\begin{itemize}
    \item Output1: is the estimated speech related to Speaker B.
    \item Output2: is the estimated speech related to Speaker A.
\end{itemize}

\vspace{0.1cm}

As shown in Fig. \ref{fig:perm}, each scenario is a possible permutation which results in alternate separation errors. For two-talker speech separation, there are two possible output-label assignments (permutation) and accordingly two possible separation errors. In general, for $S$ sources in a mixture, there are $S$! alternate possible permutations, which causes $S!$ different cost functions.  In neural network training, it is necessary to find the correct separation error (correct cost function) and then perform back-propagation through the correct cost. The general consensus is to perform backpropagation through the gradients of the minimum separation error. This is the technique used in Permutation Invariant Training (PIT) speech separation method which has been shown to be effective in addressing the permutation ambiguity \cite{yu2017permutation}. However, it has been discussed \cite{yang2020interrupted, yousefi2019probabilistic} that the hard decision on choosing the minimum cost as the best solution results in training a sub-optimal separation model. To be more specific, the process of choosing the correct separation error is more challenging in the initial epochs of training, where the network is still naive and its outputs are not reliable. In those first epochs, costs of different permutations are very similar, and minimum separation error does not necessarily represent the correct output-label assignment. Therefore, the network will be trained based on a wrong decision epoch after epoch, which finally will contribute to a sub-optimal separation model.

To address this problem, it is important to optimize both model parameters and label assignments in the training phase.  However, PIT uses a fixed label-assignment for every epoch and the benefit of using a flexible label assignment has not been explored sufficiently in the literature. The authors in \cite{yang2020interrupted} have showed that label assignments chosen based on minimum separation error may be very random especially in initial epochs where network outputs show poor results. They studied the behavior of their network in selecting the output-label assignment, and discovered that the selected label assignments for a high percentage of training examples may be reversed in two consecutive epochs \cite{yang2020interrupted}. This rapid decision flip confuses both the network and the optimizer, which leads to updating the model parameters in opposite directions. These observations are detrimental in the training phase, which manifest the inadequacy of PIT.

Additionally, in our previous study, we explored the distributions of both separation errors associated with the possible permutations in a two-talker speech separation scenario \cite{yousefi2019probabilistic}. The Kernel Distribution Estimation (KDE) of both separation error 1 and separation error 2 were plotted which revealed the inseparability of the separation errors in the first epoch of training. In the calculated KDE, error1 and error2 are more likely to be observed in regions where their values are very close \cite{yousefi2019probabilistic}. We showed that choosing the minimum cost may lead to assigning the wrong label to the network output which affects  quality  of the  trained  model. Therefore, we proposed Probabilistic PIT (Prob-PIT) \cite{yousefi2019probabilistic} in which the output-label  permutation was considered as a discrete latent random variable with a uniform prior distribution. Next, a log-likelihood function was defined based on prior distributions and separation errors of all possible permutations. We optimized the network parameters by maximizing the log-likelihood function. Unlike conventional PIT that enforces a hard decision by using one output-label permutation with the minimum cost, Prob-PIT uses all permutations by employing the soft-minimum function where leads to better overall separation results. To achieve the most optimal possible model, in he cost function of Prob-PIT, a hyperparameter was defined and manually tuned. However, manual tuning for the best value is a tedious process, which required extensive computational resources. Despite spending time and computational costs, the optimum value may not be found in challenging  situations. Therefore, in this study we define a novel training framework in which the optimum value for the cost function parameter is learned from data during the training phase.

\begin{figure*}
\begin{center}
\includegraphics[width=\linewidth]{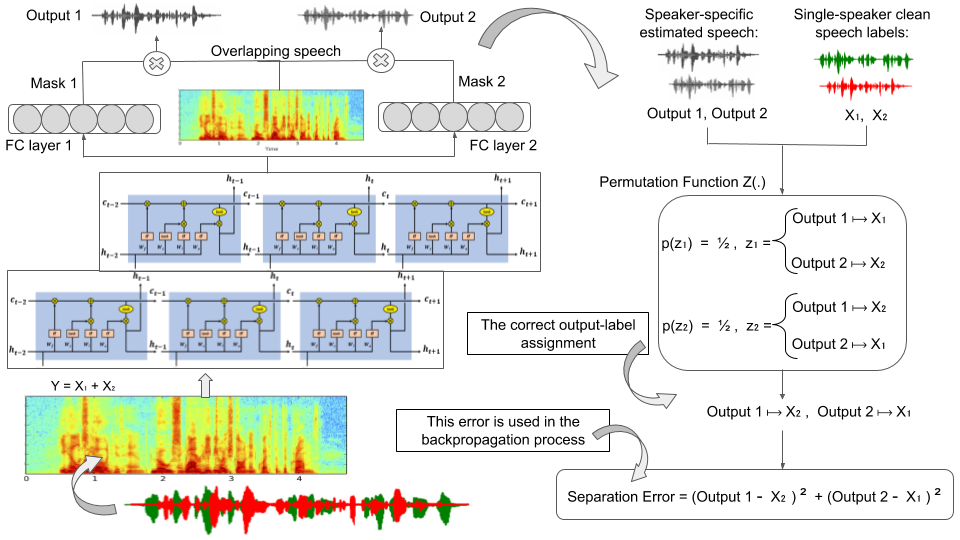}
\caption{Speech separation network architecture. First, the mixed speech is transformed to frequency domain using STFT. Next, the magnitude spectra of mixed speech is fed to a two-layer LSTM network followed by two Fully Connected (FC) layers for speaker-specific mask generation. Once the speaker-specific speech signal are estimated, they are passed to the permutation function $Z(.)$ for finding the best output-label assignment. Finally, the separation error is calculated based on the selected permutation.}
\label{fig:perm-network}
\vspace{-0.8cm}  
\end{center}
\end{figure*}

\section{Trainable Prob-PIT}
\label{sec:net}
We introduce the soft-minimum Permutation Invariant Training method in this section. As noted in Sec. \ref{sec:problem}, assume $X_1$ and $X_2$ contain magnitude spectra of speaker-specific clean speech signals shown as $X=[X_1, X_2]$ and $Y$ is the magnitude spectra of the overlapping speech as introduced in Eq. \ref{eq:2-sep}. In supervised learning, we are give pairs of mixed speech signal and their associated single-speaker speech signals in the form of $(Y,X)$ and  our task is to train a model that estimates $X$ based on the mixed speech $Y$ observation. In the following subsections, we describe the proposed generative speech model, the network architecture, and the training framework in detail.

\subsection{Model structure}
\label{sec:submodel}
In this proposed speech separation method, we define the magnitude spectra of single-speaker clean speech using a generative model as:

\begin{equation}
X = \hat{X} + \epsilon,
\label{eq:graphical_model}
\end{equation}
in which $\hat{X}$ is the estimated magnitude spectra by employing the separation system and $\epsilon$ is the separation error. For deriving $\hat{X}$, we assume the neural network $D(\theta)$ with learnable parameters $\theta$ is employed. The network $D(\theta)$ takes overlapping speech observation $Y$ as input, and estimates two speaker-specific speech signals in its outputs $O_1$ and $O_2$:
\begin{equation}
O_1, O_2 = D(Y, \theta),
\label{eq:net}
\end{equation}
Once outputs $O_1$ and $O_2$ are derived, the goal of separation is accomplished. However, to build a reliable separation model, several epochs of training are required.  In the training phase, backpropagation is performed to update the network parameters $\theta$ in order to optimize for the separation task. Backpropagation  computes the gradients of the loss function with respect to the network parameters $\theta$ for each example in the dataset (i.e. each pair of $(Y,X)$). As noted in Sec. \ref{sec:problem}, for two-talker overlapping speech separation, there are two possible permutations between outputs $O_1$ and $O_2$ with clean speech labels $X_1$ and $X_2$. These two possible permutations lead to two alternate separation costs as depicted in Fig. \ref{fig:perm-network}. This situation causes a gradient conflict because it is not clear to the optimizer which separation cost should be used for the gradient calculation in backpropagation. In contrast to PIT, we define a one-to-one permutation function $Z(.)$ to solve the permutation ambiguity. The function $Z(.)$ permutes the order of the network outputs to match the correct order of the single-speaker speech labels. In a two-talker speech separation scenario, the permutation function, $Z(.)$, can take two forms as follow: 
\begin{equation}
\hat{X} = Z(O_1, O_2),\\
\label{eq:per}
\end{equation}

As noted, the possible output-label assignment is considered as a latent variable with a uniform distribution. Therefore, in our task, the function $Z(.)$ can take two forms: $z_1(.)$ and $z_2(.)$ such that:

\begin{equation}
\begin{split}
[O_1,O_2]=z_1([O_1,O_2]),\\
[O_2,O_1]=z_2([O_1,O_2]). 
\end{split}
\end{equation}

In general, if there are $S$ active speakers in the mixture audio signal, there are $S!$ possible permutations between the network outputs and the clean speech labels. This means that the permutation function $Z(.)$ will have $S!$ possible forms, with all having the same probability of  $\frac{1}{S!}$. Only one permutation ( either $z_1$ or $z_2$) is considered as the correct response. Therefore, by replacing $\hat{X}$ in Eq. \ref{eq:graphical_model} with Eq. \ref{eq:net} and \ref{eq:per}, the generative model can be reformulated as: 

\begin{equation}
X =  Z(D(Y, \theta)) + \epsilon,
\label{eq:graphical_model1}
\end{equation}

Here, both $\epsilon$ and the permutation function $Z(.)$ are latent variables. $\epsilon$ is the estimation error which is typically modeled by a standard Gaussian distribution with mean zero and variance $\sigma^2$. Also, depending on the number of active speakers in the mixture, the permutation function $Z(.)$ could have different possible forms with uniform distribution. Eq. \ref{eq:graphical_model1} is  therefore our proposed generative model to represent single-speaker speech in this work.

\subsection{Model Training}
In Sec. \ref{sec:submodel}, we defined the proposed model pipeline. However, this model contains thousands of hyperparameters that must be trained based on extensive amounts of data. In contrast to conventional PIT, in our supervised training approach, we optimize the hyperparameters of the model by maximizing the log-likelihood function. The probability of estimating the speaker-specific speech signal $X$ conditioned on the observation of the mixed speech signal $Y$, is the likelihood of the model which we aim to maximize. According to Bayes rule, the likelihood $P(X|Y)$ can be rewritten as:

\begin{equation}
P(X|Y)=\sum_{All\ possible\ Z} P(X|Z,Y)P(Z).
\label{eqn:prob1}
\end{equation}

Here, $P(Z)$ is the probability of each possible permutation, which is set to $\frac{1}{S!}$ for $S$ active speaker in a mixed speech signal. Therefore, $P(Z)$ is not dependent on $z$. $P(X|Z,Y)$ can be derived based on the Eq. \ref{eq:graphical_model1}. The distribution of $X$ is determined by the distribution of $\epsilon$ with a new mean as:
\begin{equation}
P(X|Z,Y)=\mathcal{N}(Z(D(Y,\theta)),\,\sigma^{2}I),
\label{eqn:prob}
\end{equation}
where, $I$ is the identity matrix and $\mathcal{N}$ is the Gaussian distribution. By replacing $P(X|Z,Y)$ in Eq. \ref{eqn:prob1}, and inserting the Gaussian distribution equation, the following expression is obtained for the log-likelihood function :
\vspace{-0.05in}
\begin{equation} \label{eqn:expanded_cost}
\begin{split}
\log P(X|Y) = C + \log \sum_{All\ Z}
exp{(\dfrac{-||X-Z(D(Y,\theta))||^2}{\gamma})},
\end{split}
\vspace{-0.2in}
\end{equation}
where $\gamma=2\sigma^2$, and $C$ is a constant value that does not depend on the learnable parameters $\theta$ or the permutation latent variable $z$. However, this function does depend on the number of sources, number of frames, dimensionality of features, and variance of the estimation error. The log-likelihood function expressed in Eq. \ref{eqn:expanded_cost} is maximized in order to train the model. However, due to the logarithm function, this equation might be numerically unstable. Therefore, to ensure stability, we employ the log-sum-exp stabilization procedure: $\log \sum_i e^{x_i} = \max_i x_i + \log \sum_i e^{x_i-\max_i x_i}$. The following equations show the numerically stable form of Eq.~(\ref{eqn:expanded_cost}):

\begin{equation} \label{eqn:final}
\begin{split}
\log P(X|Y)=-e(Z_{min},\theta) + \hspace{4cm} \\
\gamma \ \log\bigg( 1 + 
\sum_{Z \neq Z_{min}} \exp\Big(
\frac{e(Z_{min},\theta)-e(Z,\theta)}{\gamma}
\Big)
\bigg),\hspace{0.2cm} \\[0.1cm]
e(Z,\theta)=||X-Z(D(Y,\theta))||^2, \hspace{2.8cm} \\[0.1cm]
Z_{min} = \arg\min_{Z} e(Z,\theta).\hspace{3.7cm} 
\end{split}
\vspace{-0.07in}
\end{equation}

Here, $e(Z,\theta)$ is the separation error of the permutation $Z$, and $Z_{min}$ is the permutation that has the minimum separation error. Noted earlier, $\gamma$ is equal to $2\sigma^2$, and $\sigma^2$ is the variance of the estimation error in the proposed model. Since $e(Z_{min})-e(z)$ is always negative, both exponential and logarithmic functions are numerically stable.

From Eq. \ref{eqn:final}, the model parameters $\theta$ are optimized to maximize the log-likelihood function $\log P(X|Y)$. This optimization is performed by applying the smooth minimum of the costs of all permutations with a smoothing factor of $\gamma$. The first part of this cost function is $-e(Z_{min},\theta)$, which is the minimum error among all possible label permutations. This is the same cost that conventional PIT uses. The second part of the equation is the cost of all other possible permutations. Therefore, if $\gamma=0$, then maximizing this cost function is equal to minimizing the PIT cost function. Alternatively, by setting $\gamma$ to larger values, a compromise is obtained between the cost of a minimum permutation versus the cost of all permutations.

\section{Experiments}
\label{sec:exp}

\subsection{Dataset}

As reported in \cite{von2019all}, real-world recording datasets such as AMI only contain approximately 5-10\% overlapping speech, which may not be sufficient for training a neural network model without overfitting the data. Therefore, we follow the same mixed speech generation process used in \cite{yu2017permutation, hershey2016deep, chen2017deep, yousefi2020frame}. We generate overlapping speech utterances based on the GRID corpus, which is a multi-speaker, sentence-based corpus used in a monaural speech separation and recognition challenge \cite{cooke2006audio}. This corpus contains 34 speakers, 16 female and 18 male speakers, each providing 1000 sentences, which have been frequently used in several overlapping speech detection and separation studies \cite{yousefi2018assessing,shokouhi2017teager,tu2015speech,yousefi2019probabilistic}.
To generate mixed speech utterances, random speech recordings are selected from random speakers. The chosen utterances are first processed through a Speech Activity Detector (SAD) for removal of silence segments. Most recordings in the GRID corpus have almost the same duration. However, for utterances with different duration, longer utterances are cut so that their length matches the shorter utterance, then they are summed with a random Signal-to-Interference Ratio (SIR), which is uniformly distributed between 0 to 5 dB. Each data sample in our generated corpus contains three waveforms, the two selected random utterances and the output generated mixed speech. For each dataset, we have generated 10h of mixed data for the training set, 4h for development, and 2h mixtures for the test set\footnote{The corpus generated here will be shared with the speech community.}. Also, speakers used for generating the test set are separate from those used in training and development sets.

\subsection{Evaluation metrics}

Speech separation techniques are usually evaluated using the blind source separation evaluation (BSS-EVAL) toolbox~\cite{vincent2006performance,wang2018supervised}. Two of the widely used measures from this toolkit are Signal-to-Distortion (SDR), Signal-to-interference Ratio (SIR), and Signal-to-Artifact Ratio (SAR) in the estimated speech signals. SIR and SAR measure different types of residual noise in the estimated speaker-specific speech signal. SIR assesses the remaining noise due to the residual interference in the separated speech signal which is called mis-separation noise. Also, SAR  measures the noise related to the reconstruction algorithm which includes, glitches due to the STFT phase estimation process. SDR measures the distortion introduced to the estimated signal by both  mis-separation and the reconstruction algorithm.

Both SDR, SIR, and SAR have been shown to be well correlated with human assessments of signal quality~\cite{fox2007modeling}. From a mathematical point of view, SDR is defined as the ratio of the target signal power to the distortion introduced by the interference, reconstruction noise, and all other background noise. SIR is defined as the ratio of the target signal power to that of the interference signal still remained in the separated speech. SAR is defined as the ratio of the estimated signal power to the reconstruction noise. In this study, we evaluate system performance using these two metrics \cite{vincent2006performance}.

\subsection{Experimental results}

For network training, a 129-dim STFT magnitude spectra is employed, computed over a frame size of 32ms with a 50\% frame shift. The network architecture consists of two Long Short-Term Memory (LSTM) Network with 128 neurons each. The output of the last LSTM layer is passed to the Softmax activation function which adds a non-linearity to the model. Next, two different fully connected layers are used to estimate speaker-specific masks. Finally, by multiplying the estimaed masks wih the overlapping speech magnitude spectra, two magnitude spectra with dimension [129 * frame-number] are generated for the two speech sources. This architecture is one of the effective structures employed by conventional PIT in \cite{yu2017permutation}. Since the memory cells in LSTM keep track of the speaker-specific information, this architecture is a suitable choice for speech separation \cite{chen2017long}. However, since a different dataset has been used in this study, we tune the hyperparameters to ensure that network has a viable initial setup for the speech separation task prior to experiments. The network is trained for 50 epochs using Adam optimization algorithm. Our  experiments  show that a dropout rate of 20$\%$, and a learning rate  of  0.0005 reduced by 0.7 when the cross-validation loss improvement is less than 0.003 in two successive epochs,  are  the  best  choices  for  the  hyperparameters.

\begin{table*}
\caption{The results of the proposed Soft-minimum PIT speech separation in terms of Signal to Distortion ratio (SDR), Signal to Interference Ration (SIR), and Signal to Artifacts Ratio (SAR) for both speakers in the mixture. $Gamma=2$ maximizes the log-likelihood of he separation cost and results in the best performance.}
\setlength{\extrarowheight}{4pt}
\setlength{\tabcolsep}{20pt}
\makebox[\textwidth][c]{
\begin{tabular}{cccccccc}
\toprule
&      \multicolumn{3}{c}{Speaker 1} & \multicolumn{3}{c}{Speaker 2} \\
\cmidrule(lr){2-4}\cmidrule(lr){5-7}

Constant Gamma & SDR  & SIR & SAR  & SDR & SIR & SAR \\ \midrule
Gamma 0 & 6.6693 & 8.1966 & 13.0914 & 2.9977 &	4.5250 & 10.3076\\ 
Gamma 1 & 6.8775 & 8.4905 & 13.0699 & 3.2571 &	4.8916 & 10.2669\\
\textbf{Gamma 2} & \textbf{7.1894} & \textbf{8.9429} & \textbf{13.0547} & \textbf{3.6061} &	\textbf{5.3652} & \textbf{10.2844}\\
Gamma 3 & 6.9817 & 8.6556 &	13.0409 & 3.3853 &	5.0869 & 10.2152\\
Gamma 4 & 6.8852 & 8.4423 &	13.4668 & 3.1657 &	4.6821 & 10.9357\\
\bottomrule
\end{tabular}
}
\label{tab:const}
\end{table*}

\textbf{Baseline PIT--} In this study, we compare our proposed method with PIT introduced in \cite{yu2017permutation}. PIT minimizes the mean-square-error of the estimated speech signal to train the network. To do so, PIT selects the permutation with a minimum error throughout backpropagation. The cost function used in PIT is as:

\begin{equation}
    Cost_{(PIT)} =||X-\hat X||^2 = ||X-Z_{min}(D(Y,\theta))||^2.
\end{equation}

However, in our proposed method, we train the network by maximizing the log-likelihood of all separation errors introduced in Eq. \ref{eqn:final} as:

\begin{equation}
\begin{split}
\log P(X|Y)= -e(Z_{min},\theta) + , \hspace{6.3cm} \\
 \gamma \ \log\bigg( 1 + 
\sum_{Z \neq Z_{min}} \exp\Big(
\frac{e(Z_{min},\theta)-e(Z,\theta)}{\gamma}
\Big)
\bigg)\hspace{4.3cm} \\ 
\end{split}   
\label{eq:part}
\end{equation}
Here, if $\gamma$ is set to zero, then the second term drops. Therefore, in the case of $\gamma=0$, our proposed cost function is:

\begin{equation}
   \log P(X|Y) = -e(Z_{min},\theta) = - ||X-Z_{min}(D(Y,\theta))||^2
\end{equation}
which is the same cost function used in conventional PIT. Therefore, in the case of $\gamma=0$, maximizing the log-likelihood of our proposed cost function is equal to minimizing the mean-square-error used in PIT. Hence, $\gamma=0$, which is the same as PIT approach is considered as baseline in this study.

\textbf{Constant Gamma Soft-minimum PIT--} The results of our proposed approach is shown in Table \ref{tab:const}. Evaluation metrics are reported for both speaker-specific estimated speech signals. For each reconstructed waveform, SDR, SIR, and SAR are reported.  As mentioned in the previous section, $Gamma=0$ is considered as the baseline. Larger values of Gamma provides more weight to the second term of the proposed cost function in Eq. \ref{eq:part}. Therefore, with larger values of Gamma, the optimization process focuses more weight on cost of all possible permutations rather than the minimum cost. Therefore, Gamma is  an important parameter that should be selected carefully. As shown in Table \ref{tab:const}, $1\leq Gamma \leq 5$ improves the PIT baseline, however, the choice of $Gamma=2$ seems to be the best choice with the best output performance in terms of SDR and SIR. Larger choices of $Gamma$ result in under-performing PIT, which is expected as the optimizer ignores the permutation with the minimum cost. 

Additionally, as mentioned in Sec. \ref{sec:net}, $Gamma$ is equal to $\gamma=2*\sigma ^2$, in which $\sigma$ is the variance of the separation error. Therefore, when $Gamma=2$, the variance of the separation error is set to $\sigma=0.25$ which is a reasonable choice for the separation cost.

It is worth mentioning that, the core role of $Gamma$ is in the training phase. Our motivation in defining the log-likelihood function in Eq. \ref{eqn:final} was to prevent the network parameters to be trained based on an unreliable cost function. Therefore, the log-likelihood function uses $Gamma$ to replace the minimum cost by the soft-minimum cost function which results in a smoother optimization landscape and therefore the model is less likely to converge to a poor local minimum. Once the model is trained, the separation performance is evaluated by other metrics such as SDR, SIR, and SAR. Thus, in the testing phase, we follow the PIT testing procedure and  set $\gamma$ to zero. 

\begin{figure*}
\centering
\begin{tabular}{cccc}
\hspace{-2mm}
\includegraphics[width=8cm]{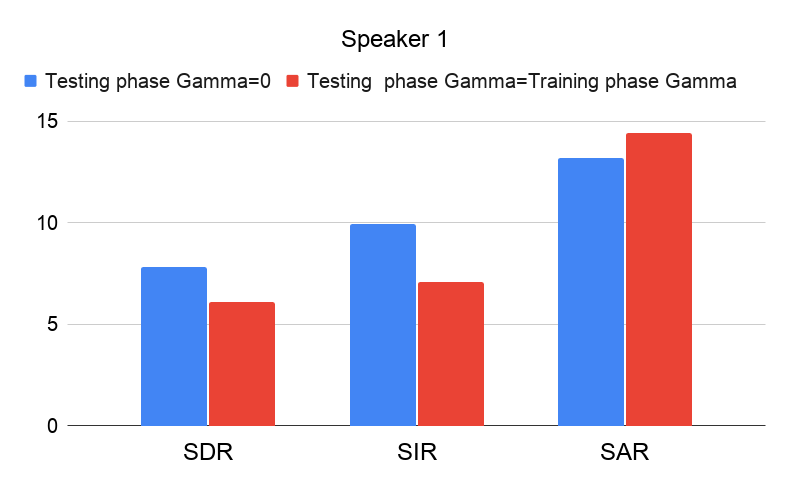}&
\hspace{1cm}
\includegraphics[width=8cm]{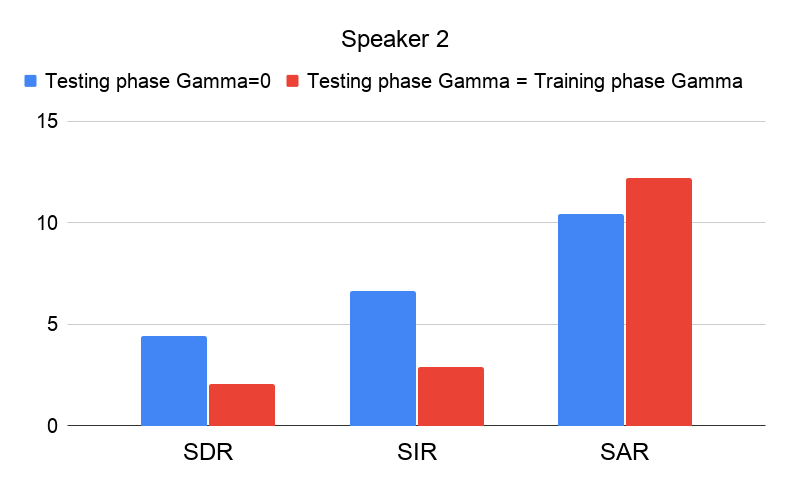}&
\hspace{-2mm}
\end{tabular}
\vspace{-0.1in}
\caption{The comparison of speech separation performance for different Gamma values in the testing phase. As depicted in the plots, once the model is trained with the Soft-minimum PIT method, Gamma should be set to Zero in the testing phase. This is because the trained model is reliable in the testing phase and the permutation with the minimum loss is the correct output-label assignment.  }
%\vspace{-4mm}
\label{fig:test}
\end{figure*}

During testing, we still need to find the correct output-label assignment for determining SDR, SIR, and SAR. In our proposed approach, there are two possible options in the testing phase: \emph{(i)} choosing the output-label assignment with minimum separation loss, and \emph{(ii)} choosing the output-label assignment with the permutation that maximizes the log-likelihood function with the same assigned $Gamma$ in the training phase. These two approaches are evaluated and plotted in Fig. \ref{fig:test}. Here, the blue bars represent the first case in which $Gamma$ is set to zero during the testing phase. Alternatively, the red bars represent the case in which $Gamma$ in the testing phase is the same value as in the training set. As depicted in the plot, system performance is much better when $Gamma=0$ in terms of SDR, SIR, and SAR. This is expected as our proposed method is aimed at modifying the training process to result in a higher quality separation model with improved performance. As mentioned in section \ref{sec:problem}, choosing the permutation with a minimum cost in the training phase confuses the model and optimizer, and leads to updating the model parameters toward opposite directions in successive epochs. However, once the model is trained by maximizing the log-likelihood of all possible separation costs, the testing phase should be performed by selecting the permutation with the minimum cost.

\begin{table*}
\caption{The results of the proposed Soft-minimum PIT with trained Gamma in terms of Signal to Distortion ratio (SDR), Signal to Interference Ration (SIR), and Signal to Artifacts Ratio (SAR) for both speakers in the mixture. The initial value of $Gamma=1$ is the best option for learning its optimal value.}
\setlength{\extrarowheight}{4pt}
\setlength{\tabcolsep}{20pt}
\makebox[\textwidth][c]{
\begin{tabular}{cccccccc}
\toprule
&      \multicolumn{3}{c}{Speaker 1} & \multicolumn{3}{c}{Speaker 2} \\
\cmidrule(lr){2-4}\cmidrule(lr){5-7}

Trained Gamma & SDR  & SIR & SAR  & SDR & SIR & SAR \\ \midrule
\textbf{Gamma 1} & \textbf{7.6471} &	\textbf{9.8187} &	\textbf{12.6322} &	\textbf{4.2793} &	\textbf{6.6073} &	\textbf{9.6374} \\
Gamma 2 & 7.2199 &	9.0586 &	12.8717 &	3.6579 &	5.5009 &	10.1288 \\ 
Gamma 3 & 7.1882 &	8.9946 &	12.8792 &	3.6487 &	5.4935 &	10.0569 \\
Gamma 4 & 7.0982 &	8.8117 &	13.1499 &	3.4787 &	5.1941 &	10.4676 \\
\bottomrule
\end{tabular}
}
\label{tab:trained}
\end{table*}

\textbf{Trainable Gamma Soft-minimum PIT--} The main limitation of the soft-minimum PIT is finding the right value for $Gamma$. This can be exhausting since it requires time and computational resources which makes it a sub-optimal tuning process. Therefore, instead of tuning this parameter, in this section, $Gamma$ is trained by the optimizer. To do so, the gradients of the log-likelihood cost function should be calculated with respect to $Gamma$ as well. 

In Eq. \ref{eqn:expanded_cost}, the constant $C$ contains $\gamma$, however, since in our previous experiments, $\gamma$ was considered as a constant factor which did not affect the gradients, and therefore we ignored the constant $C$. However, in this set of experiments, $\gamma$ is considered as a learnable parameter and its gradients affect the optimizer. Therefore, Eq. \ref{eqn:expanded_cost} is modified as:

\begin{equation} \label{eqn:p}
\begin{split}
\log P(X|Y) = C^{'}+ \dfrac{1}{\gamma} + \log \sum_{All\ Z}
exp{(\dfrac{-||X-Z(D(Y,\theta))||^2}{\gamma})},
\end{split}
\end{equation}
In the modified cost function, the constant $C^{'}$ does not depend on any trainable parameter, so its gradient is zero in the backpropagation. It is worth noting that, since $\gamma$ is in the denominator, it has the potential to unstablize the cost function. Therefore, it is important to add $\epsilon$ to $\gamma$ in the denominator to assure the stability of the training process. The separation performance for the trained $Gamma$ is reported in Table \ref{tab:trained}. The reported values for $Gamma$ in the first column are the initial values for this trainable parameter. As reported in the table, the initial value of $Gamma=1$ leads to the best performance in terms of SDR and SIR. In our experiments, after several epochs, $Gamma$ converged to a number in the range $[1.5\leq \gamma \leq 2]$, and it was very close to the manually selected value reported in the previous table. Since the optimal value of $\gamma$ for the data used is in the range of $[1,2]$, then setting the initial value of $\gamma$ higher than $1$ makes it more challenging for the optimizer to converge to the optimal $\gamma$, causing an overall lower performance.  
\begin{figure*}
\centering

\begin{tabular}{ccccc}
%\hspace{-7mm}
\includegraphics[width=8cm,height=5.5cm]{ 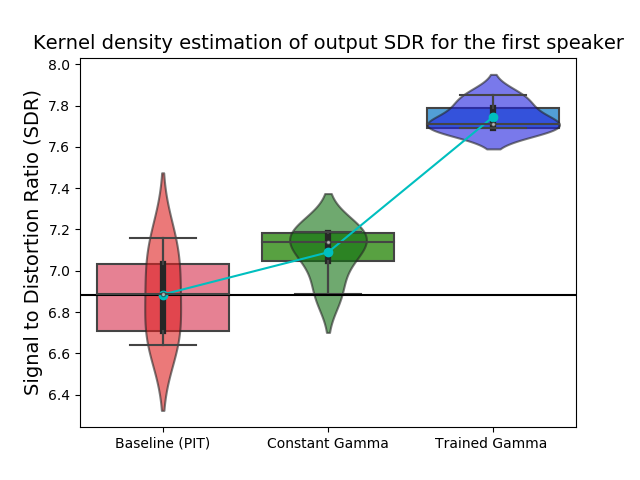}&
%\hspace{-7mm}
\includegraphics[width=8cm,height=5.5cm]{ 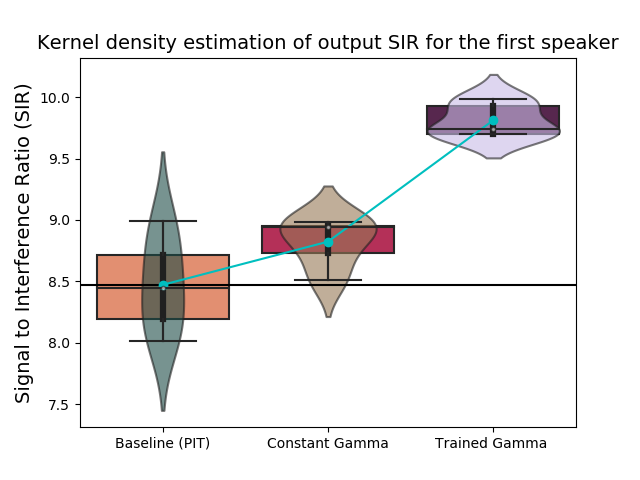}\\
%\hspace{12mm}
\includegraphics[width=8cm,height=5.5cm]{ 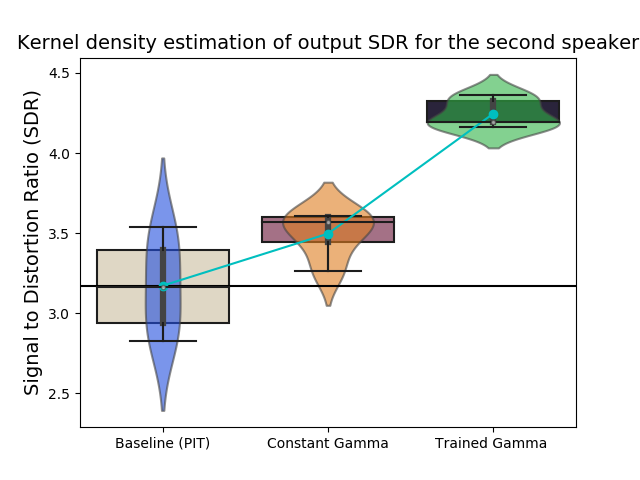}&
%\hspace{-2mm}
\includegraphics[width=8cm,height=5.5cm]{ 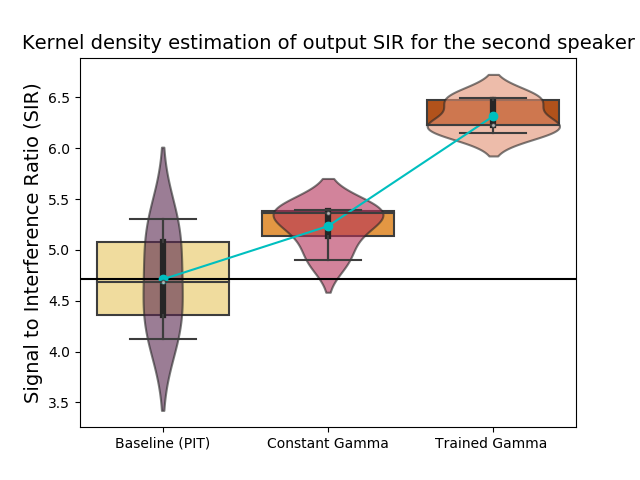}
	
\end{tabular}
%\vspace{-0.1in}
\label{figur}\caption{The boxplot and kernel distribution of the Baseline (PIT), Soft-minimum PIT with Constant Gamma, and Soft-minimum PIT with trained Gamma is depicted. For each separation system, 5 experiments have been performed. Each violin-shaped object represents the boxplot and the kernel distribution estimation of those five experiments. The black solid line represents the mean of the results for the PIT baseline. The blue circles on the blue line are the mean of the output evaluation metrics for other two separation method: Constant Gamma in Soft-minimum PIT and Trained Gamma in the Soft-minimum PIT. As shown in the figure, the proposed soft-minimum PIT in both scenarios outperforms PIT baseline. Additionally, the output SDR and SIR of the proposed method have a lower variance for both separated speakers.}
\vspace{-4mm}
\label{fig:violin}
\end{figure*}

Furthermore, in order to minimize the effect of parameter initialization on our final separation metrics, we train each network five times with different initial parameters. The distributions of the final evaluation metrics of those five experiments are depicted in Fig. \ref{fig:violin}. In these plots, the boxplot and kernel distribution estimation of those five experiments are depicted. The boxplot is a standardized way of displaying the distribution of experiments based on a five number summary, which are the minimum, first quartile (Q1), median, third quartile (Q3), and maximum.
Additionally, each violin-shaped object represents the kernel distribution estimation of the results. The more the result points are in a specific range, the larger the violin is for that range. Also, each blue circle on the blue line represents the mean of the evaluation metrics in the five experiments for each separation system. The black solid line is the mean of the performance metrics for $\gamma=0$ as the PIT baseline.  

The first row in Fig. \ref{fig:violin} represents the distribution estimation for the fist separated speaker in terms of SDR (first row, first column), and SIR (first row, second column). Likewise, the distribution estimation plots are depicted for the second speaker in the second row. As shown, the PIT separation method has a long and narrow violin response with high variance of the results. This system behavior is not desired because the output performance is not reliable and may tend to vary significantly. However, in our proposed soft-minimum PIT for both constant $Gamma$ and Trained $Gamma$ the violin plots tend to be wide and short which confirms a small variance in the output results. However, as mentioned before, since manual tuning of $Gamma$ does not guarantee finding the optimal value, this approach is sub-optimal compared to learning $Gamma$ in the training phase. Once the optimal value for $Gamma$ is learned by the optimizer, the SDR of the first speaker is improved by almost +1dB compared to the baseline PIT. This is almost +13\% relative improvement with lower variance, revealing the superiority and reliability of the proposed approach compared to PIT. In terms of SIR, training $Gamma$ in the proposed soft-minimum PIT achieves the best results with +1.5dB improvement compared to PIT. Similar to SDR, the variance of the output SIR is lower in the Trained $Gamma$ scenario. Nevertheless, as depicted in the second row of Fig. \ref{fig:violin}, the pattern of the results is repeated for the second speaker as well. The output SDR and SIR for the second speaker have the same relative improvement in the proposed approach compared to the baseline. This is a very important accomplishment, because the second speaker has lower energy in the mixture (due to the lower input SDR we chose in the mixing process), therefore, recovering such a degraded signal is very challenging. Additionally, most separation systems are only effective in recovering the target speaker speech from the mixture at the expense of the remaining speaker-specific speech signals. Therefore, an effective speech separation solution should be capable of recovering both speakers from the mixture with the same level of quality. Similar to the first speaker, once soft-minimum PIT is employed in the training phase, the variance of the output SDR and SIR for the second speaker is also lowered compared to PIT.

\section{discussion}
\label{sec:dis}

In general, speech separation is usually performed in two steps \emph{(i)} separating the specific speech sources, and \emph{(ii)} determining the best output-label assignment,  allowing for assessment of separation error via the evaluation metrics. The second step called \emph{label permutation ambiguity} has been a long standing challenge in training neural networks for speech separation. Recently proposed \emph{Permutation Invariant Training (PIT)} addressed this problem by determining the output-label assignment which minimizes the separation error. In PIT, a neural network is trained that separates the speaker-specific speech signals. In the training phase, PIT determines the best output-label assignment which minimizes the separation error. Next, backpropagation is performed based on the minimum separation error. However, studies \cite{yousefi2019probabilistic, yang2020interrupted} have shown that choosing the minimum separation error is a hard decision imposed on the optimizer, especially in the initial epochs of training where network is still naive. Each possible output-label assignment result in a different cost function. In the initial epochs of the training, the value of these costs are very close. Therefore, the minimum separation error does not necessarily represent the correct output-label assignment. Additionally, in the beginning of the training phase, the selected label assignments may be reversed in two consecutive epochs which confirms the unreliability of the network output. If backpropagation is performed based on only the minimum separation error, then the rapid decision flip confuses both the network and optimizer, which leads to updating the model parameters toward opposite directions. Therefore, updating network parameters based on the cost of one single permutation is not an optimal solution, and leads to an inefficient training of the network. These observations are detrimental in the training phase, which manifests the inadequacy of PIT.

In contrast to PIT, we propose the Soft-minimum PIT which considers the output-label assignment as a latent variable with uniform distribution. In Soft-minimum PIT, the network is trained by maximizing the log-likelihood of the prior distributions and the separation errors of all possible permutations. Since in the proposed method, all possible output-label assignment are taken into consideration in the backpropagation, the optimization landscape becomes smoother, which is in contrast to the hard decision of minimizing the Mean-Square-Error of the minimum separation cost performed in PIT. In the Soft-minimum PIT, the smoothness of the cost function is controlled by $\gamma$, which is $2*\sigma^2$ with $\sigma$ being the variance of the separation error. In this work, we have explored both tuning and training $\gamma$ in the proposed method to evaluate the separation performance. The results of our experiments on the simulated two-talker overlapping speech dataset shows that Soft-minimum PIT outperforms PIT significantly (p-value $ < 0.01$). Also, the greatest improvement is achieved by training $\gamma$ with other parameters of the network. Trained $\gamma$ in the Soft-minimum PIT results in improved output SDR and SIR by +1dB and +1.5dB with lower variance during multiple repeated experimental runs with different initialization.  

The effectiveness of the proposed Soft-minimum PIT can be attributed to several reasons. The core strength of Soft-minimum PIT is the incorporation of all possible output-label assignments in training the model parameters. This is in contrast to PIT, which uses a hard decision in assigning the output-label permutation that minimizes the total separation error. During training, the network is not able to estimate the speaker-specific speech signals correctly, therefore, its decision in assigning the correct output-label permutation is not reliable. Also, since in the initial epochs of  training, all  model parameters are randomly selected, so the separated speech signals at the output are far from the desired speech signals.  Consequently, the separation error of different possible permutations are very similar and the correct output-label assignment does not necessarily have a minimum separation loss. Therefore, if the selected output-label permutation in PIT is not correct, then the model parameters are updated based on a wrong decision, resulting in deteriorating the training process.

In addition, it has been shown \cite{yang2020interrupted} that the output-label assignment selected in PIT tends to change in two successive epochs for most of the data samples in the corpus. This uncertainty in finding the correct permutation causes a disorientation in the optimizer because the model parameters are updated in opposite directions for most of the initial epochs. Hence, in our Soft-minimum PIT, we consider the costs of all possible permutations for training the network in a probabilistic framework. 

Another reason for the success of our proposed approach is that the minimum cost function used in PIT is replaced by a soft-minimum function. In several applications of machine learning, it has been shown that replacing the minimum by the soft-minimum results in a smoother optimization landscape and therefore it is less likely to converge to a poor local minima. This can also be explained in terms of the decision flips that PIT experiences during training. Since in Soft-minimum PIT the decisions are reliable and comprehensive, then the optimization landscape does not have many poor local minimums.
Two core observations in this study confirms this finding for speech separation as well. First, SDR and SIR values of the soft-minimum are better than PIT significantly (p-value $ < 0.01$); (2) the variance of  SDR and SIR values are lower for both constant and trained $\gamma$. A lower variance in the results show a more stable system, which may be caused by a smoother optimization landscape.

\section{Conclusion}
\label{sec:conc}

In this study, we proposed Soft-minimum PIT to address label permutation ambiguity in speech separation. For Training single-channel speaker-independent speech separation models, two steps are required: first, estimating the speaker-specific speech signal; second, finding the correct output-label assignment for calculating the separation error. The second step known as label permutation ambiguity has been a long-standing challenge in training neural networks for the task of speech separation. One general solution introduced in PIT proposes to train a neural network based on the output-label assignment with minimum separation cost. Unfortunately, the hard choice of minimum cost permutation is not the best technique, especially in initial epochs of training where the network is still not strong enough to effectively separate the speech signals. In contrast to PIT, in our proposed Soft-minimum PIT, we consider all possible permutations as a discrete latent variable with a uniform prior distribution. Next, we trained the network by maximizing the log-likelihood function defined based on  prior distributions and  separation errors of all possible permutations. In our proposed approach the smoothness of the decision was controlled by a variable parameter that can be either tuned or trained. In this study, we explored both cases and results based on GRID datasets show that the proposed Soft-minimum PIT significantly outperforms PIT in terms of SDR and SIR. This solution therefore offers a viable option to effectively separate overlap/mixed speaker audio streams, especially in naturalistic audio scenarios.

% or


% \appendices

% \section{}

% \section*{Acknowledgment}


\ifCLASSOPTIONcaptionsoff
  \newpage
\fi



%\bibliographystyle{IEEEtran}
%\begin{thebibliography}{1}
%\end{thebibliography}


\bibliographystyle{IEEEbib}
\bibliography{refs}

% \end{IEEEbiography}


% % insert where needed to balance the two columns on the last page with
% %\newpage



%\vfill


\end{document}